\documentclass[submission,copyright,creativecommons]{eptcs}
\usepackage{tikz}
\usepackage{alltt}
\usepackage{bussproofs}
\providecommand{\event}{PxTP 2019} 
\usepackage{breakurl}             
\usepackage{underscore}           
\usepackage{amsthm}
\usepackage{hyperref}
\usepackage{listings}
\usepackage{enumitem}

\usepackage{xcolor}

\theoremstyle{definition}

\tikzstyle{lien}=[->,>=stealth,rounded corners=5pt,thick]

\title{\textsc{ekstrakto} \\A tool to reconstruct \textit{Dedukti} proofs from
TSTP files \\(extended abstract)}
\author{Mohamed Yacine El Haddad
  \institute{CNRS}
\and
Guillaume Burel
\institute{ENSIIE}
\and
Fr\'ed\'eric Blanqui
\institute{INRIA}
\and
\footnotesize LSV, CNRS, ENS Paris-Saclay, Universit\'e Paris-Saclay
}
\def\titlerunning{\textsc{ekstrakto}, A tool to reconstruct \textit{Dedukti}
proofs from TSTP files}
\def\authorrunning{M. Y. El Haddad \& G. Burel \& F. Blanqui}
\begin{document}
\maketitle

\begin{abstract}
    Proof assistants often call automated theorem provers to prove subgoals.
    However, each prover has its own proof calculus and the proof traces that
    it produces often lack many details to build a complete proof. Hence these
    traces are hard to check and reuse in proof assistants. \textsc{Dedukti} is a proof
    checker whose proofs can be translated to various proof assistants: Coq,
    HOL, Lean, Matita, PVS. We implemented a tool that extracts TPTP
    subproblems from a TSTP file and reconstructs complete proofs in \textsc{Dedukti}
    using automated provers able to generate \textsc{Dedukti} proofs like ZenonModulo or
    ArchSAT.
    This tool is generic: it assumes
    nothing about the proof calculus of the prover producing the trace, and
    it can use different provers to produce the \textsc{Dedukti} proof. We applied
    our tool on traces produced by automated theorem provers on
    the CNF problems of the TPTP library and we were able to reconstruct a
    proof for a large proportion of them, significantly increasing the number of
    \textsc{Dedukti} proofs that could be obtained for those problems.
\end{abstract}

\section{Introduction}
In order to discharge more burden from the users of interactive
theorem provers, and thus to widen the use of these tools, it is
crucial to automate them more. To achieve this goal, in
the process of checking the validity of formulas, proof assistants
could use an external theorem prover to automate their tasks and obtain a
proof of a specific formula. Once a proof is found, the proof assistant
applies this proof on the current goal and tells the user that all is
done in background.  However, this can work only if the prover builds a
complete proof that is easily checkable by the proof assistant. We
distinguish two families of automated theorem provers: some provers, like \textit{ZenonModulo}~\cite{zenonmodulo}
and \textit{ArchSAT}~\cite{buryarchsat}, output
complete proofs but are not very efficient to find a proof; others, like \textit{E prover}~\cite{Schulz:LPAR-2013}
and \textit{ZipperPosition}~\cite{zipperposition}, are
more powerful but return only proof traces, i.e.\ proofs with less
details.

In this paper we are interested in first-order automated theorem provers which can return TSTP~\cite{Sut17} traces. We
will use \textsc{Dedukti}~\cite{dedukti} as proof checker because
\textsc{Dedukti} files can be translated to many other proof assistants (Coq, HOL, Lean Matita, PVS)~\cite{thire18sharing}.

We start by presenting the \textsc{TPTP/TSTP} formats with an example. Then, we
describe how proofs and formulas are encoded in \textsc{Dedukti}. We then present
our solution implemented in a tool named \textsc{ekstrakto} in two steps:
extraction of sub-problems and proof reconstruction. Finally, we conclude and
give some perspectives.

\section{TPTP}
TPTP~\cite{Sut17} is a standard library of problems to test automated theorem
provers~\cite{casc}.
Each TPTP file represents a problem in propositional, first-order or higher-order logic. We distinguish the type of formulas by using one of the
keywords:
CNF, FOF, TFF and THF,
corresponding respectively to mono-sorted first-order formulas in
clausal normal form, general mono-sorted first-order formulas, typed
first-order formulas, and typed higher-order formulas.

In this work, we restrict our attention to CNF formulas since their proofs use logical consequences only,
which is not the case of FOF formulas (e.g. Skolemisation).

Apart from an include instruction, each line in a TPTP file is a declaration of
a formula given with its role, e.g.\ axiom, hypothesis, definition or conjecture:

\begin{alltt}
    cnf(name, role, formula, information).
\end{alltt}

TSTP~\cite{Sut17} is a library of solutions to TPTP problems. In this paper, we
call a TSTP file a trace. It is obtained after running an automated theorem prover
on a TPTP problem. The syntax used in a TSTP file is the same as TPTP except for
the content of the \textit{information} field. This field contains general
information about how the current formula is obtained. Here is the grammar used
to describe a source in the \textit{information} field:
\begin{alltt}
<source>               :== <dag_source> | [ <sources> ] | ...
<dag_source>           :== <name> | inference(..., ..., <inference_parents>)
<inference_parents>    :== [] | [ <sources> ]
<sources>              :== <source> (, <source>)*
\end{alltt}
For our purpose only 3 cases are of interest as shown in the grammar above:

\begin{enumerate}[label=\arabic*)]
    \item When it is the name of a formula previously declared.
    \item When it is a list of several sources:
\begin{alltt}
    [s_0, s_1, ..., s_n]
\end{alltt}
    \item When it is an inference:
\begin{alltt}
    inference(name, infos, [s_0, s_1, \ldots, s_n])
\end{alltt}

\end{enumerate}

The name of the inference refers to the name of the rule used by the prover to
prove the current step. The \textit{infos} field contains more information about
the inference like status, inference name, etc.
Note that each $ s_i $ is a source and therefore can contain sub-inferences.

Here is an example of a TSTP file obtained after running
\textit{E prover} on the TPTP problem SET001-1:

\begin{lstlisting}[frame=single, title=SET001-1.p]
cnf(c_0,axiom,
    ( subset(X1,X2)
    | ~ equal_sets(X1,X2) )).
cnf(c_1,hypothesis,
    ( equal_sets(b,bb) )).
cnf(c_2,axiom,
    ( member(X1,X3)
    | ~ member(X1,X2)
    | ~ subset(X2,X3) )).
cnf(c_3,negated_conjecture,
    ( ~ member(element_of_b,bb) )).
cnf(c_4,hypothesis,
    ( member(element_of_b,b) )).
cnf(c_5,hypothesis,
    ( subset(b,bb) ),
    @inference(spm,[status(thm)],[c_0,c_1])@).
cnf(c_6,hypothesis,
    ( member(X1,bb)
    | ~ member(X1,b) ),
    @inference(spm,[status(thm)],[c_2,c_5])@).
cnf(c_7,negated_conjecture,
    ( $false ),
    @inference(cn,[status(thm)],[inference(rw,[status(thm)],
    [inference(spm,[status(thm)],[c_3,c_6]),c_4])])@,
    [proof]).
\end{lstlisting}

We can represent this trace as the following tree:

\begin{prooftree}\hspace*{-1cm}
        \AxiomC{$ \vdash \texttt{Form(c\_3)} $}
            \AxiomC{$ \vdash \texttt{Form(c\_2)} $}
                \AxiomC{$ \vdash \texttt{Form(c\_0)} $}
                \AxiomC{$ \vdash \texttt{Form(c\_1)} $}
            \RightLabel{spm}
            \BinaryInfC{$ \vdash \texttt{Form(c\_5)} $}
        \RightLabel{spm}
        \BinaryInfC{$ \vdash \texttt{Form(c\_6)} $}
        \RightLabel{spm}
        \BinaryInfC{}
        \AxiomC{$ \vdash \texttt{Form(c\_4)} $}
    \RightLabel{rw}
    \BinaryInfC{}
    \RightLabel{cn}
    \UnaryInfC{$ \vdash \texttt{Form(c\_7)} $}
\end{prooftree}

where:
\begin{verbatim}
Form(c_0) = subset(X1,X2) | ~equal_sets(X1,X2)
Form(c_1) = equal_sets(b,bb)
Form(c_2) = member(X1,X3) | ~member(X1,X2) | ~subset(X2,X3)
Form(c_3) = ~member(element_of_b,bb)
Form(c_4) = member(element_of_b,b)
Form(c_5) = subset(b,bb)
Form(c_6) = member(X1,bb) | ~member(X1,b)
Form(c_7) = $false
\end{verbatim}

\section{First-order logic in \textsc{Dedukti}}\label{sec:fofded}

\textsc{Dedukti} is a proof checker based on the $\lambda\Pi$-calculus
modulo rewriting~\cite{dedukti}. In \textsc{Dedukti}, one can declare (dependent) types and
function symbols, and rewriting rules for defining these symbols. We describe
how a formula and its proof are encoded  in \textsc{Dedukti} using the
\textit{Curry-Howard} correspondence, i.e., we interpret
formulas as types and their proofs as terms. In the following, we
recall the encoding of first-order logic in \textsc{Dedukti}, as
described in~\cite{dedukti}. This encoding is used in
\textit{ZenonModulo}, which is an extension to
rewriting of the automated theorem prover \textit{Zenon}~\cite{zenon}. \textit{ZenonModulo} outputs \textsc{Dedukti} files after having found a proof using the tableaux method. The following file defines the type of sorts, the type of terms, the type of formulas and then the type of proofs.

\vfill
\begin{lstlisting}[frame=single, title=zen.lp, mathescape=true]
symbol sort : TYPE          // Dedukti type for sorts
symbol $\iota$ : sort             // default sort

symbol term : sort $\Rightarrow$ TYPE  // Dedukti type for sorted terms

symbol prop : TYPE          // Dedukti type for formulas
symbol $\dot{\bot}$ : prop
symbol $\dot{\top}$ : prop
symbol $\dot{\neg}$ : prop $\Rightarrow$ prop
symbol $\dot{\land}$ : prop $\Rightarrow$ prop $\Rightarrow$ prop
symbol $\dot{\lor}$ : prop $\Rightarrow$ prop $\Rightarrow$ prop
symbol $\dot{\Rightarrow}$ : prop $\Rightarrow$ prop $\Rightarrow$ prop
symbol $\dot{\forall}$ : $\forall$ a, (term a $\Rightarrow$ prop) $\Rightarrow$ prop
symbol $\dot{\exists}$ : $\forall$ a, (term a $\Rightarrow$ prop) $\Rightarrow$ prop
symbol $\dot{=}$ : $\forall$ a, term a $\Rightarrow$ term a $\Rightarrow$ prop

symbol Proof : prop $\Rightarrow$ TYPE  // interprets formulas as types
rule Proof ($\dot{\Rightarrow}$ &a &b) $\rightarrow$ Proof &a $\Rightarrow$ Proof &b
// rewriting rule defining the type of proofs for $\dot{\Rightarrow}$
\end{lstlisting}

Now, for each TSTP file, we generate a Dedukti file defining its signature by declaring a Dedukti symbol $f$ for each function symbol $f$ of the TSTP file:

\begin{lstlisting}[frame=single, title=SET001-1.lp, mathescape=true]
symbol element_of_b : zen.term $\iota$
symbol subset       : zen.term $\iota$ $\Rightarrow$ zen.term $\iota$ $\Rightarrow$  zen.prop
symbol b            : zen.term $\iota$
symbol member       : zen.term $\iota$ $\Rightarrow$ zen.term $\iota$ $\Rightarrow$  zen.prop
symbol bb           : zen.term $\iota$
symbol equal_sets   : zen.term $\iota$ $\Rightarrow$ zen.term $\iota$ $\Rightarrow$  zen.prop
\end{lstlisting}

Hence, every formula of first-order logic can be represented in \textsc{Dedukti} by
using the function $\varphi$ defined as follows:

\begin{center}

\begin{tabular}{rcl}
    $ \varphi(x)  $ & := & $  x $ \\
    $ \varphi(f (t_1, t_2, \dots, t_n))$ & := & $ f ~ \varphi(t_1) ~ \varphi(t_2) ~ \ldots ~ \varphi(t_n) $ \\
    $ \varphi (\bot) $ & := & $ \dot{\bot} $ \\
    $ \varphi (\top) $ & := & $ \dot{\top} $ \\
    $ \varphi(\neg A) $ & := & $ \dot{\neg} \varphi (A) $ \\
    $ \varphi (A \land B) $ & := & $ \varphi(A)  \ \dot{\land}  \ \varphi (B) $ \\
    $ \varphi (A \lor B) $ & := & $ \varphi(A)  \ \dot{\lor}  \ \varphi (B) $ \\
    $ \varphi (A \Rightarrow B) $ & := & $ \varphi(A)\,\dot{\Rightarrow}\,\varphi (B) $ \\
    $ \varphi (\forall x A) $ & := & $ \dot{\forall}_\iota(\lambda x,\varphi(A)) $ \\
    $ \varphi (\exists x A) $ & := & $ \dot{\exists}_\iota (\lambda x,\varphi(A)) $ \\
    $ \varphi (x = y) $ & := & $ \varphi(x) \ \dot{=}_\iota \ \varphi(y) $ \\
\end{tabular}
\end{center}

For example,
\[
    \varphi(\forall X_1, \forall X_2, s (X_1, X_2) \lor \neg e (X_1,
    X_2)) :=
    \dot{\forall}_\iota (\lambda X_1,\dot{\forall}_\iota (\lambda X_2,
    (s ~ X_1 ~ X_2) ~ \dot{\lor} ~ \dot{\neg} (e ~ X_1 ~ X_2)))
\]

For every formula $A$, its proof in \textsc{Dedukti} is a term that has the
type $Proof (\varphi(A))$. One can define a similar embedding for proofs as the
one we presented for first-order formulas, as shown in~\cite{dedukti}.

\section{Architecture}
In this section, we explain in details how \textsc{ekstrakto} works.
In order to produce a \textsc{Dedukti} proof from a TSTP file, \textsc{ekstrakto}
extracts a TPTP problem for each formula declaration containing at least one
inference, and calls \textit{ZenonModulo} (or any other automated prover producing
\textsc{Dedukti} proofs, see discussion below) on each generated
problem to get a \textsc{Dedukti} proof for this problem. If the external prover
succeeds to find a proof of all the generated problems, then we combine
those proofs in another \textsc{Dedukti} file to get a \textsc{Dedukti} proof of
the whole TSTP file.

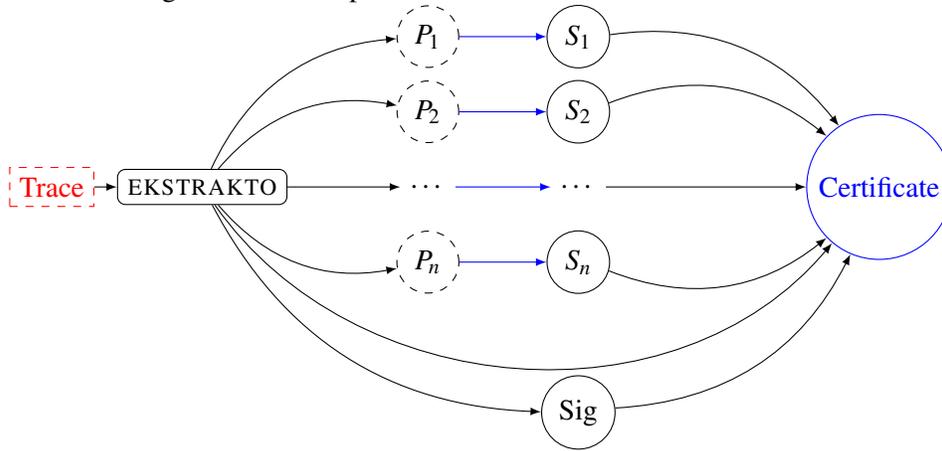
\begin{tikzpicture}
    \node[draw, color=red, dashed] (Trace)                   at (0,0) {Trace};

    \node[draw, rectangle, rounded corners=3pt] (Ekstrakto)  at (2,0) {\textsc{ekstrakto}};

    \node[draw, circle, dashed] (P1)                         at (5,2) {$P_1$};
    \node[draw, circle, dashed] (P2)                         at (5,1) {$P_2$};
    \node                       (Points)                     at (5,0) {\ldots};
    \node[draw, circle, dashed] (Pn)                         at (5,-1){$P_n$};

    \node[draw, circle]         (S1)                         at (7,2) {$S_1$};
    \node[draw, circle]         (S2)                         at (7,1) {$S_2$};
    \node                       (SPoints)                    at (7,0) {\ldots};
    \node[draw, circle]         (Sn)                         at (7,-1){$S_n$};

    \node[draw, circle]         (Signature)                  at (7,-3){Sig};

    \node[draw, color=blue, circle]         (Certificate)    at (11,0) {Certificate};

    \draw[->,>=latex] (Trace) edge (Ekstrakto);

    \draw[->,>=latex] (Ekstrakto) edge [bend left] (P1);
    \draw[->,>=latex] (Ekstrakto) edge [bend left] (P2);
    \draw[->,>=latex] (Ekstrakto) edge (Points);
    \draw[->,>=latex] (Ekstrakto) edge [bend right](Pn);
    \draw[->,>=latex] (Ekstrakto) edge [bend right](Signature);
    \draw[->,>=latex] (Ekstrakto) edge [out=-55,in=-130] (Certificate);

    \draw[->,>=latex, color=blue] (P1) edge (S1);
    \draw[->,>=latex, color=blue] (P2) edge (S2);
    \draw[->,>=latex, color=blue] (Points) edge (SPoints);
    \draw[->,>=latex, color=blue] (Pn) edge (Sn);

    \draw[->,>=latex] (S1) edge [bend left] (Certificate);
    \draw[->,>=latex] (S2) edge [bend left] (Certificate);
    \draw[->,>=latex] (SPoints) edge (Certificate);
    \draw[->,>=latex] (Sn) edge [bend right](Certificate);
    \draw[->,>=latex] (Signature) edge [bend right](Certificate);

\end{tikzpicture}

\subsection{Extracting TPTP problems}

To extract a TPTP problem from a trace step, we need to find the premises used in it. We define the function $ \mathcal P $
which takes a TSTP source as input and returns the set of premises used by the prover:
\[ \mathcal P (name) = \{name\} \]
\[ \mathcal P ([s_0, s_1, \ldots, s_n]) = \bigcup_{i=0}^n \mathcal P (s_i) \]
\[ \mathcal P (inference(name, infos, [s_0, s_1, \ldots, s_n])) = \bigcup_{i=0}^n \mathcal P
(s_i) \]

Note that if we have an inference $ t $ inside another one, say $ s $, we will
repeat the process for each sub-inference and omit $ s $ from the set of
premises, i.e., if we represent an inference step by a proof tree we take only
the leaves of this tree as premises.

We omit all information that is not needed (\textit{status},
\textit{name}, \ldots). In particular we do not consider the inference name field.
Even if it could be used to fine-tune the problem, we prefer to ignore it in
order to remain generic since the names are specific to the prover that produced
the trace. Hence, we have:

\[
    \mathcal P (inference([inference([inference([c\_3, c\_6]), c\_4]) ])) =
    \{c\_3, c\_6, c\_4 \}
\]

After getting all the premises used for proving $\verb|Form|(name)$, say $name_0,\ldots,name_k$, we generate the following TPTP problem:

\[
    \verb|Form|(name_0) \Rightarrow \ldots \Rightarrow \verb|Form|(name_k) \Rightarrow \verb|Form|(name)
\]

Note that the generated TPTP problem is a FOF formula. The reason
of this choice is to keep the same formula when we combine the sub-proofs. If
we generated a CNF problem, then we would need to negate the goal and
it would be more complex to reconstruct the proof.

Since we are using FOF formulas in sub-problems that are obtained from
a CNF trace, we need to quantify over each free variable to get a
closed formula.

In our example, there are 3 steps (colored in blue in the file \verb|SET001-1.p| above). \textsc{ekstrakto} will generate the following 3 first-order formulas:
\begin{alltt}
    \(\verb|Form(c_0)| \Rightarrow \verb|Form(c_1)| \Rightarrow \verb|Form(c_5)| \) \\
    \(\verb|Form(c_2)| \Rightarrow \verb|Form(c_5)| \Rightarrow \verb|Form(c_6)| \) \\
    \(\verb|Form(c_3)| \Rightarrow \verb|Form(c_6)| \Rightarrow \verb|Form(c_4)| \Rightarrow \verb|Form(c_7)| \) \\
\end{alltt}

Each formula will be written in a separate TPTP file as follows:

\begin{lstlisting}[frame=single, title=c_5.p]
    fof(c_5, conjecture, (
            (![X1, X2] : (s (X1, X2)|~equal_sets (X1, X2)))
        =>  ((equal_sets (b, bb))
        =>  (subset (b, bb))))).
\end{lstlisting}

\begin{lstlisting}[frame=single, title=c_6.p]
    fof(c_6, conjecture,(
            (![X1, X2, X3] : (member (X1, X3)|~member (X1, X2)
                              |~subset (X2, X3)))
        =>  ((subset (b, bb))
        =>  (![X1] : (member (X1, bb)|~member (X1, b)))))).
\end{lstlisting}

\begin{lstlisting}[frame=single, title=c_7.p]
    fof(c_7, conjecture, (
            (~member (element_of_b, bb))
        =>  ((![X1] : (member (X1, bb)|~member (X1, b)))
        =>  ((member (element_of_b, b))
        =>  ($false))))).
\end{lstlisting}

\subsection{Proof reconstruction}

If the automated theorem prover succeeds to solve all the generated TPTP
problems, then we can reconstruct a proof in \textsc{Dedukti} directly by using
the proof tree of the trace that we are trying to certify and all the proofs of
the sub-problems. The proof term of each sub-problem is irrelevant since it has
the right type.

The global proof is reconstructed from each sub-proof. We just need to apply
each proof term of a sub-proof to its premises by following the proof tree
of the TSTP file. Indeed, the type of the sub-proof of
$\verb|Form|(name)$ using premises $name_0,\ldots,name_k$ is
\begin{lstlisting}[mathescape=true]
zen.Proof ($\dot{\Rightarrow}$ $\varphi$(Form($name_0$)) ($\dot{\Rightarrow}$ $\varphi$(Form($name_1$)) ...
                               ($\dot{\Rightarrow}$ $\varphi$(Form($name_k$)) $\varphi$(Form($name$)))$\dots$))
\end{lstlisting}
Thanks to the rule given in zen.lp in Section \ref{sec:fofded}, this
type is convertible to
\begin{lstlisting}[mathescape=true]
zen.Proof ($\varphi$(Form($name_0$))) $\Rightarrow$ ... $\Rightarrow$ zen.Proof ($\varphi$(Form($name_k$))) $\Rightarrow$ zen.Proof ($\varphi$(Form($name$)))
\end{lstlisting}
Hence, the proof term of a sub-problem is a function whose arguments are
proofs of the premises and which returns a proof of its conclusion.
Since we are handling only CNF formulas, the proof that we want to
reconstruct at the end is always a proof of $ \bot $.
Before applying those proof terms we need to declare our hypotheses. With our example file we get:

\begin{lstlisting}[frame=single, title=proof_SET001-1.lp, mathescape=true]
definition proof_trace
 (hyp_c_0 : zen.Proof ($\varphi$(Form(c_0))))
 (hyp_c_1 : zen.Proof ($\varphi$(Form(c_1))))
 (hyp_c_2 : zen.Proof ($\varphi$(Form(c_2))))
 (hyp_c_3 : zen.Proof ($\varphi$(Form(c_3))))
 (hyp_c_4 : zen.Proof ($\varphi$(Form(c_4))))
  : zen.Proof $\dot{\bot}$
  :=
  let lemma_c_5 = c_5.delta hyp_c_0 hyp_c_1           in
  let lemma_c_6 = c_6.delta hyp_c_2 lemma_c_5         in
  let lemma_c_7 = c_7.delta hyp_c_3 lemma_c_6 hyp_c_4 in
  lemma_c_7
\end{lstlisting}

where \textit{delta} is the name of the proof term in each file.

All this has been implemented in a tool called \textsc{ekstrakto}\footnote{https://github.com/elhaddadyacine/ekstrakto} consisting of 2,000 lines of OCaml.

\section{Experiments}

    We run the \textit{E prover} (version 2.1) on the set of CNF problems
of TPTP library v7.2.0 (7922 files) with 2GB of memory space and a
timeout of 5 minutes. We obtained 4582 TSTP files.
 On these TSTP files, \textsc{ekstrakto} generated 362556 TPTP files.
 \textit{ZenonModulo} generated a \textsc{Dedukti} proof for 90\% of these files,
 \textit{ArchSAT} generated 96\% and the union of both produced 97\% \textsc{Dedukti} proofs:

\begin{table}[h]
    \caption{Percentage of \textsc{Dedukti} proofs on the 362556 extracted TPTP files}
    \centering
    \begin{tabular}{ |c|c| }
        \hline
        Prover & \% TPTP \\
        \hline
        \textit{ZenonModulo}                         &  90\%  \\
        \textit{ArchSAT}                             &  96\%  \\
        \hline
        \textit{ZenonModulo} $\cup$ \textit{ArchSAT} &  97\%   \\
        \hline
    \end{tabular}
\end{table}
However, as it suffices that no \textsc{Dedukti} proof is found for only one TPTP file for
getting no global proof, \textsc{ekstrakto} can generate a complete proof
for only 48\% of TSTP files using \textit{ZenonModulo}, 61\% using \textit{ArchSAT}
and 72\% using at least one of them:
\begin{table}[h]
    \caption{Percentage of \textsc{Dedukti} proofs on the 4582 TSTP files generated by \textit{E prover}}
    \centering
    \begin{tabular}{ |c|c| }
        \hline
        Prover & \% TSTP\\
        \hline
        \textit{ZenonModulo}                         &  48\% \\
        \textit{ArchSAT}                             &  61\% \\
        \hline
        \textit{ZenonModulo} $\cup$ \textit{ArchSAT} &  71\% \\
        \hline
        \end{tabular}
    \end{table}

Consequently, we are now able to produce 2189 \textsc{Dedukti} proofs from the
TPTP library using \textit{E prover} and \textit{Zenon Modulo} (resp. 2793 using
\textit{E prover} and \textit{ArchSAT} and 3285 using \textit{E prover},
\textit{Zenon Modulo} and \textit{ArchSAT}), whereas under the same
conditions, \textit{Zenon Modulo} alone is only able to produce 1026 \textsc{Dedukti}
proofs (resp. 500 for \textit{ArchSAT} alone).

    Sometimes, \textit{ZenonModulo} and \textit{ArchSAT} fail to find a
    proof even if the sub-problem is simpler than the main one.
    This is justified by the fact that the proof calculus used
    in \textit{ZenonModulo} and \textit{ArchSAT} is based on a
    different method from the one used in
    \textit{E prover}. In fact, some steps that are trivial for a prover
    based on resolution or superposition may not be trivial for
    \textit{ZenonModulo} or \textit{ArchSAT} which use the tableaux method.

    \textit{iProverModulo} is another candidate to prove TSTP steps,
    but it performs some transformations before outputting a \textsc{Dedukti} proof. Therefore the proof reconstruction is hard in the sense that we need to justify each
    transformation made by \textit{iProverModulo}.

\section{Conclusion and perspectives}
We have presented a tool that reconstructs proofs generated by first-order
theorem provers. We described how proofs and formulas are represented in
\textsc{Dedukti} and how we can implement a simple proof reconstruction.

The advantage of \textsc{ekstrakto} is to be generic since
it does not depend on the rules used by the automated prover to find the proof.
Another advantage is the fact that the proofs are expressed in \textsc{Dedukti},
i.e., we can translate them to many other
systems (Coq, HOL, Lean, Matita, PVS).

In our experiments, we used \textit{ZenonModulo} and \textit{ArchSAT} to prove each trace
step since they are tools that produce \textsc{Dedukti} proof terms.

\textsc{ekstrakto} should be extended to handle non-provable steps like Skolemisation.
This latter technique could possibly be implemented using the method described
in~\cite{dowek}. We should also be more generic, by supporting more features of TSTP like typed
formulas and  definitions introduced by the prover.

\bigskip
{\bf Acknowledgements.} The authors thank the anonymous reviewers for
their useful comments. This research was partially supported by the
Labex DigiCosme (ANR11LABEX0045DIGICOSME) operated by ANR as part of
the program "Investissement d'Avenir" Idex ParisSaclay
(ANR11IDEX000302).

\bibliographystyle{eptcs}
\bibliography{main}

\end{document}